\documentclass[a4paper,preprintnumbers,showpacs,onecolumn,superscriptaddress,nofootinbib,amsmath,amssymb,notitlepage]{revtex4-2}
\usepackage{mathtools,leftindex,tensor,mhchem}
\usepackage{booktabs}
\usepackage{siunitx}
\usepackage{enumitem}
\usepackage{times}
\usepackage{dcolumn}
\usepackage{mathrsfs}
\usepackage{comment}
\usepackage{hyperref}
\usepackage[utf8]{inputenc}
\usepackage{amsmath,accents}
\usepackage{mathtools}
\usepackage{bbm}
\usepackage{bm}
\usepackage{amsfonts}
\usepackage{amssymb}
\usepackage{mathrsfs}
\usepackage[scr=rsfs,cal=boondox]{mathalfa}
\usepackage{wasysym}
\usepackage{graphicx}
\usepackage[]{subfigure}
\usepackage{filecontents}
\usepackage[dvipsnames]{xcolor}
\usepackage{bbold}
\usepackage[scr=rsfs,cal=boondox]{mathalfa}

\usepackage{stackengine}
\stackMath
\newcommand\tsup[2][2]{%
 \def\useanchorwidth{T}%
  \ifnum#1>1%
    \stackon[-1.3ex]{\tsup[\numexpr#1-1\relax]{#2}}{\mathchar"307E}%
  \else%
    \stackon[-1ex]{#2}{\mathchar"307E}%
  \fi%
}
\hypersetup{
    bookmarks=true,         
    pdftoolbar=true,        
    pdfmenubar=true,        
    pdffitwindow=true,     
    pdfstartview={FitH},     
    pdftitle={My title},     
    pdfauthor={author},      
    pdfsubject={Subject},    
    pdfcreator={Creator},    
    pdfproducer={Producer},  
    pdfkeywords={keyword1} 
    pdfnewwindow=true,       
    colorlinks=true ,        
    linkcolor=Blue  ,        
    citecolor=Blue,      
    filecolor=Blue,       
    urlcolor=Blue            
}

\newcommand{\arctanh}{\operatorname{arctanh}}

\newcommand{\ed}{\mathrm{d}}

\newcommand{\cn}{\operatorname{\mathbf{cn}}}
\newcommand{\nc}{\operatorname{\mathbf{nc}}}

\newcommand{\m}{\mathcal{k}}

\newcommand{\oalpha}[1]{\accentset{\circ}{\alpha}}
\newcommand{\obf}[1]{\accentset{\circ}{\mathbf{f}}}
\newcommand{\boR}[1]{\accentset{\circ}{\mathbf{R}}}
\newcommand{\obF}[1]{\accentset{\circ}{\mathbf{F}}}
\newcommand{\obPi}[1]{\accentset{\circ}{\mathbf{\Pi}}}

\usepackage{scalerel}
\usepackage{tikz}
\usetikzlibrary{svg.path}

\definecolor{orcidlogocol}{HTML}{A6CE39}
\tikzset{
  orcidlogo/.pic={
    \fill[orcidlogocol] svg{M256,128c0,70.7-57.3,128-128,128C57.3,256,0,198.7,0,128C0,57.3,57.3,0,128,0C198.7,0,256,57.3,256,128z};
    \fill[white] svg{M86.3,186.2H70.9V79.1h15.4v48.4V186.2z}
                 svg{M108.9,79.1h41.6c39.6,0,57,28.3,57,53.6c0,27.5-21.5,53.6-56.8,53.6h-41.8V79.1z M124.3,172.4h24.5c34.9,0,42.9-26.5,42.9-39.7c0-21.5-13.7-39.7-43.7-39.7h-23.7V172.4z}
                 svg{M88.7,56.8c0,5.5-4.5,10.1-10.1,10.1c-5.6,0-10.1-4.6-10.1-10.1c0-5.6,4.5-10.1,10.1-10.1C84.2,46.7,88.7,51.3,88.7,56.8z};
  }
}

\newcommand\orcidicon[1]{\href{https://orcid.org/#1}{\mbox{\scalerel*{
\begin{tikzpicture}[yscale=-1,transform shape]
\pic{orcidlogo};
\end{tikzpicture}
}{|}}}}
\begin{document}


\title{Interacting Generalized Chaplygin-Jacobi gas: Thermodynamics approach
%
}
\author{Gilberto Aguilar-Pérez\orcidicon{0000-0001-6821-4564}}
\email{gilaguilar@uv.mx}
\affiliation{Facultad de F\'{\i}sica, Universidad Veracruzana 91097, Xalapa, Veracruz, M\'exico\\
}

\author{Miguel Cruz
\orcidicon{0000-0003-3826-1321}}
\email{miguelcruz02@uv.mx}
\affiliation{Facultad de F\'{\i}sica, Universidad Veracruzana 91097, Xalapa, Veracruz, M\'exico\\
}

\author{Mohsen Fathi\orcidicon{0000-0002-1602-0722}}
\email{mohsen.fathi@ucentral.cl}
\affiliation{Centro de Investigaci\'{o}n en Ciencias del Espacio y F\'{i}sica Te\'{o}rica (CICEF), Universidad Central de Chile, La Serena 1710164, Chile\\
}

\author{J. R. Villanueva\orcidicon{0000-0002-6726-492X}}
\email{jose.villanueva@uv.cl}
\affiliation{Instituto de Física y Astronomía, Universidad de Valparaíso, Gran Bretaña 1111, Valparaíso, Chile}


\begin{abstract}
This work investigates a cosmological model featuring an interaction between dark energy and dark matter, where the dark energy component is described by the Generalized Chaplygin-Jacobi gas (GCJG). In this study, we establish a system in which the GCJG and a pressureless dark matter fluid exchange energy via a linear interaction term, $Q \propto \rho_x$, being $\rho_{x}$ the dark energy density. By solving the conservation equations, we derive analytical expressions for the evolution of the dark energy and dark matter densities. The thermodynamic properties of this interacting system are then thoroughly analyzed. The thermodynamic analysis reveals that both dark components maintain positive temperatures, ensuring stability. Notably, the dark energy component transitions to a phantom regime in the past, a feature of interest for recent cosmological observations, without violating thermodynamic principles. The total entropy production is shown to be in agreement with the second law of thermodynamics. Furthermore, an analysis of the specific heats suggests that while the dark matter sector remains thermodynamically stable, the dark energy sector undergoes a late-time phase transition, consistent with its entering into the phantom domain at effective level.
\bigskip

{\noindent{\textit{keywords}}: dark energy, gravitation, cosmology, thermodynamics}\\

\end{abstract}

\maketitle

\tableofcontents

\section{Introduction and Motivation}\label{sec:intro}

In recent decades, cosmological models have undergone significant developments to more accurately describe the dynamics and composition of the universe. Observations of the cosmic microwave background, the accelerated expansion of the universe, and large-scale structure formation have led to the establishment of the $\Lambda$CDM model as the standard cosmological paradigm. This model posits that approximately 27\% of the universe's energy content consists of cold dark matter (CDM), while about 68\% is made up of dark energy, typically represented by a cosmological constant ($\Lambda$), which drives the accelerated cosmic expansion.

Although it fits well with each data set when examined separately, combined analyzes reveal an interconnected set of tensions that involve the Hubble constant, CMB lensing, spatial curvature, neutrino masses, and dark energy properties \cite{DiValentino:2026uua}. In this sense, the $\Lambda$CDM model faces several theoretical challenges, including the cosmic coincidence problem and the unknown nature of dark matter and dark energy. As a result, extensions of the standard model have emerged, introducing the possibility of a direct interaction between these two dark components. Such interacting dark energy models allow for an exchange of energy and/or momentum between dark energy and dark matter, modifying the universe’s evolution in potentially observable ways.
These models have garnered considerable attention, as they may help alleviate current cosmological tensions—such as the Hubble tension—and provide insights into the fundamental physics of the dark sector. Moreover, the inclusion of interaction terms can influence the dynamics of structure formation, the growth of cosmic perturbations, and the evolution of galaxy clusters.

As mentioned before, models positing an interaction between the dark energy and dark matter sectors were originally introduced to address the cosmic coincidence problem, as the energy exchange can create attractor solutions that keep the components' energy densities of the same order of magnitude. A primary virtue of these models is their ability to alter both the cosmic expansion history and the effective gravitational constant governing structure formation. However, their main defect is the lack of motivation for first principles, which has led to a plethora of kernels of phenomenological interactions, some of which can introduce theoretical instabilities or are penalized by model selection criteria for their added complexity. Despite these challenges, current observational constraints consistently allow for a small, non-zero coupling. Several recent analyzes suggest a preference for an interaction in which dark energy decays into dark matter (e.g., a positive coupling parameter for kernels proportional to an energy density). This direction of energy flow is not only favored by thermodynamic arguments, but is also capable of alleviating the $H_{0}$ tension by reducing the Universe's expansion rate at high redshifts compared to the concordance model; for an interesting review on interaction models, see, for instance, \cite{Wang_2016, Wang_2024, vanderwesthuizen2025ilinearinteractingdark}. A new study offers strong statistical evidence for interactions between dark matter and dark energy, challenging the validity of the standard cosmological model. Using recent datasets such as DESI and Planck, the authors argue that these exchange mechanisms account for the observed anomalies in the universe’s expansion more effectively than conventional explanations. The analysis considers two different theoretical approaches—coupled quintessence and coupled fluid—and in both cases finds a preference for interaction, with a significance reaching up to $5\sigma$. These results indicate that the observed discrepancies are not caused by time-varying dark energy alone, but rather by a genuine physical coupling between the components of the dark sector \cite{li2026strongevidencedarksector}. 

Linear models typically define the interaction kernel as being directly proportional to the density of dark matter, dark energy, or a simple sum of the two. A key issue with many linear models is their susceptibility to pathologies; for instance, scenarios where energy flows from dark matter to dark energy often lead to unphysical negative energy densities in the past or future. In contrast, non-linear models often define the interaction as proportional to the product of the densities \cite{CRUZ2019114623, Aguilar-Pérez_2022}, a form that naturally vanishes when either component's density approaches zero. This feature provides a significant theoretical advantage, as it inherently prevents the energy densities from becoming negative. This distinction also affects how the models address the coincidence problem; while certain linear models can solve the problem by creating a stable, constant ratio between the dark components, non-linear models typically only alleviate it, as the interaction's influence fades in the asymptotic past and future when one component becomes completely dominant. Furthermore, their predictions for the universe's ultimate fate can diverge; in the phantom regime, a future big rip singularity is often inevitable for linear models, whereas some non-linear models can avoid this fate if the interaction remains dominant and counteracts the phantom energy's effect \cite{vanderwesthuizen2025iinonlinearinteractingdark}. As we shall see in the following, all these characteristics are discussed within our model.

This line of research continues to expand, incorporating both phenomenological approaches and motivations from fundamental theories, with the aim of offering a more comprehensive and coherent description of the universe's past, present, and future. In pursuit of understanding the dark side of the universe, modern cosmology has sought models to explain accelerated expansion. The most auspicious scenarios involve dynamical fields, where one of the proposals is the Chaplygin gas (CG) \cite{Kamenshchik_2001} model to serve as an  alternative to the cosmological constant. The CG model involves using an exotic equation of state (EoS):
$p = -\frac{B}{\rho}$, where $p$ is the pressure and $\rho$ is the energy density. In this way, one can model the transition from a matter dominated universe to one going through an accelerated expansion \cite{Kamenshchik_2001}.
The model can be extended to the Generalized Chaplygin gas model (GCG) \cite{Bento_2002} with the introduction of an additional parameter, namely $\alpha$, resulting in an EoS:
$p = -\frac{B}{\rho ^\alpha }$. See, for example, \cite{Zhang_2006}, where an analogy between  the Chaplygin gas and the interaction scheme was found.
Looking at the \textit{coincidence} problem, it is usually assumed that Dark Energy ($x$) is coupled with Dark Matter ($m$). In the standard cosmological model, dark matter and dark energy are assumed to interact solely through gravity. However, more general scenarios permit a direct transfer of energy between these two sectors, producing late-time effects that can imitate a dynamical dark energy component or even phantom-like behavior. In such models, the usual conservation equations for each component are altered by introducing an interaction term that enables energy exchange. The interaction is represented by the introduction of a new term \(Q \) called the \textit{interaction kernel}:
\begin{eqnarray}\label{eq:densmat}
    \dot \rho_{m} + 3H(\rho_{m}+p_m) = Q, \\
    \label{eq:densgcj}
     \dot \rho_{x} + 3H(\rho_{x}+p_x) = -Q.
\end{eqnarray}
In this first installment we will consider that 
the energy transfer is proportional to the dark energy density, i. e. $Q = 3Hb^{2}\rho_x$  \cite{suwa12,delCampo:2008jx}.
One reason for choosing this form is that it keeps the equations relatively simple and allows us to obtain analytical expressions for the evolution of the cosmological quantities. In addition, this type of term guarantees that energy is transferred from the dark energy sector to dark matter, a flow direction consistent with the second law of thermodynamics and with recent observational findings that indicate a preference for dark energy decay, as discussed above. Also, this interaction naturally becomes very small at early times, when the dark energy density is still small compared to the dark matter density. Because of these reasons, this interaction term provides a convenient and commonly used framework to study the thermodynamic behavior of the interacting GCJG model.

This work is organized as follows: Section \ref{sec:overview} reviews the fundamental properties and formalism of the GCJG model. In Section \ref{sec:EHTconst.}, we construct the interacting cosmological model, in which the GCJG, representing the dark energy sector, exchanges energy with the matter sector through a linear interaction term ($Q\propto\rho_{x}$). We then derive the exact analytical expressions for the evolution of the dark energy and dark matter densities. Section \ref{sec:thermo} provides a detailed analysis of the thermodynamic properties of this interacting system, demonstrating that both dark components maintain positive temperatures, which ensures thermodynamic stability and guaranties the fulfillment of the second law of thermodynamics. Finally, Section \ref{sec:conclusions} presents our summary and concluding remarks. Throughout this work, we adopt natural units where $G=c=1=M_{\rm{pl}}$. The sign convention is $(- + + \,+)$, and the primes denote differentiation with respect to the radial coordinate.

\section{The Generalized Chaplygin Jacobi dark fluid}\label{sec:overview}

Considering the fundamental nature of the CG and its generalizations, they emerge as powerful elements in describing the dynamics of the universe across its various evolutionary stages \cite{Bili__2002}. However, when confronted with observational data, the original CG reveals inconsistencies that prompt the search for further generalizations. Beyond the GCG, the GCJG, proposed in Ref. \cite{Rengo_15}, presents an alternative model in the context of inflationary cosmology. This model employs the Hamilton-Jacobi approach with the generating Hubble function
\begin{equation}\label{hcj}
    H(\bar{\phi}, \m) = H_0\, {\nc}^{\frac{1}{1+\alpha}}\Bigl([1+\alpha]\Phi \Bigr),
\end{equation}
where $\bar{\phi}$ is the scalar \textit{inflaton} field, $\Phi=\sqrt{6\pi/m_\mathrm{P}^2}\left(\bar{\phi}-\bar{\phi}_0\right)$ is a dimensionless quantity with $m_{\mathrm{P}}$ being the Planck mass, $H_0\equiv H(\bar{\phi}_0,\m)$, and  ${\nc}(x)= 1/{\cn}(x)$, with ${\cn}(x)
\equiv \cn(x; \m)$ being the Jacobi elliptic
cosine function with argument $x$ and modulus $\m$\footnote{The incomplete elliptic integral of the first kind is given by the general form  {\cite{handbookElliptic}}
\begin{eqnarray}\nonumber
    w={\bf{F}}(u,\m)=\int_0^u\frac{\ed t}{\sqrt{1-\m^2\sin^2{t}}},
\end{eqnarray}
where $u={\bf{F}}^{-1}(w,\m)={\bf{am}}(w,\m)$ is the Jacobi amplitude. This way, the Jacobi elliptic cosine function is defined as
\begin{eqnarray}\nonumber
\cos{u}=\cos{\big({\bf{am}}(w,\m)\big)=\cn(w,\m)}.
\end{eqnarray}}. 
Thus, the generating function \eqref{hcj} 
yields the EoS
\begin{equation}
\label{eq:prro2}
p=-\frac{B\,\m}{\rho^{\alpha}}-2\m'\rho
+\frac{\m'}{B}\rho^{\alpha+2},
\end{equation}
where $B$ is a real constant and $\m'=1-\m$ is the complementary modulus of the elliptic function. 
Note that for $\m=1$ (or $\m'=0$), $B>0$
and  $\alpha=1$ the CG is recovered,
whereas the GCG is obtained for $\alpha \neq 1$.
For $\m \neq 1$ the GCJG 
presents great versatility depending on 
the parameters $(B, \m)$, allowing 
for a different comparison with observational data 
and making the theory more reliable in explaining 
the accelerated expansion of the universe. While the modulus $\m$ controls the deviation from the standard GCG model, the parameter $\alpha$ determines the non-linear scaling of the pressure with density, and $B$ sets the characteristic energy scale of the fluid. As we will demonstrate below, our analysis shows that, within the range of $\alpha$ and $B$ under consideration, the system does not exhibit thermodynamic pathologies, such as negative temperatures or violations of entropy production. Beginning from the next section, we explore the properties and consequences of the interaction between the components of the dark sector, where dark energy corresponds to the GCJG and dark matter is a pressure-less fluid.


\section{Interacting GCJDF}\label{sec:EHTconst.}

\subsection{Dark Energy Density}

Since we are interested in the versatility offered by the GCJG model, we test its properties under the interacting scheme adopting the DE-DM decay mediated by the linear interaction term given as $Q = 3 H b^2 \rho_x$, where $b^{2}$ is a dimensionless coupling parameter that determines both the magnitude and the sign of the interaction, given that $b^{2}>0$. An interaction term of the form specified above solves the $H_{0}$ tension, as discussed in \cite{Di_Valentino_2017, Di_Valentino_2020, Di_Valentino_20201}. As shown in \cite{M_B_Gavela_2009}, negative values of the coupling constant aggravate the cosmological coincidence problem. Additionally, without specifying any particular model, in Refs. \cite{M_B_Gavela_2009, Gavela_2010} was found the origin of early-time, non-adiabatic, large-scale instabilities that appear in many coupled models with $\omega_{x}$, where $\omega_{x}$ denotes the dark energy equation of state in the absence of dark coupling. These instabilities are linked to the coupling terms that modify the propagation of dark energy pressure waves through the dark matter background. The size and sign of the dark-coupling terms are essential, as they determine which parameter combinations correspond to stable and unstable regimes, according to the following expression
\begin{equation}
    \Delta = \frac{Q}{3H\rho_{x}(1+\omega_{x})},
\end{equation}
when $\Delta$ is positive, the terms that depend on the dark coupling can dominate the behavior of dark energy perturbations, causing them to enter a runaway regime characterized by unstable growth. Notice that, when we adopt the linear interaction term introduced above, we are constrained by the stability requirement $1+\omega_{x} < 0$. In our framework, the equation of state of uncoupled dark energy is defined as $\omega_{x} = p_{x}/\rho_{x}$, where the dark energy pressure is given by equation (\ref{eq:prro2}). Under a suitable choice of the model parameters, condition $1+\omega_{x} < 0$ is fully satisfied. Due to the properties of the model with positive interaction term, in Ref. \cite{Di_Valentino_20201} this case is referred to as the {\it coupled phantom model}. As we show below, our model aligns with all these earlier findings in the literature.


According to these assumptions, Eq. (\ref{eq:densgcj}) may be expressed as
\begin{equation}\label{eqa1}
    \frac{{\rm d} \rho_{x}}{{\rm d} \eta} + (1+b^2) \rho_{x} + p_x  = 0,
\end{equation}
where $\eta \equiv \ln a^3$. Treating GCJG as the dark energy constituent, $\rho_x=\rho_{cj}$, Eq. (\ref{eqa1}) yields
\begin{equation}\label{eqa2}
\ln{\left(\frac{a_0}{a}\right)}^{3}=
\int_{\rho_x}^{\rho_{x0}}
    \frac{\rho_x^\alpha\, {\rm d}\rho_x}{F(\rho_x)},
\end{equation}where the function
$F$ is given by
\begin{equation}
    \label{eqa3}
    F(\rho_x)=\m  B+[2\m'-(1+b^2)] \rho_x^{1+\alpha}
    -\frac{\m'}{B}\rho_x^{2(1+\alpha)}.
\end{equation}
Performing the variable change $u = \rho_{x}^{\alpha+1}$, Eq. (\ref{eqa2}) is reduced to
\begin{equation}
    \label{eqa4}
    \rho_x(a, 1)=\left[\frac{B}{1+b^2}+\frac{(1+b^2)\rho_{x0}^{1+\alpha}-B}{1+b^2}
   \left(\frac{a_0}{a}\right)^{3(1+\alpha)(1+b^2)}\right]^{\frac{1}{1+\alpha}},
\end{equation}for $\m=1$, while for 
$0\leq \m < 1$ the integration yields
\begin{equation}\label{eqa5}
  \ln\left(\frac{a_0}{a}\right)^ {3\beta(1+\alpha) }=2\arctanh\left[\frac{2\m'\,u-
  B(2\m'-1-b^2)}{\beta\, B}\right],
\end{equation}where
\begin{equation}
    \label{beta1}\beta=\sqrt{(1+b^2)^2 -4\m' b^2}.
\end{equation}
Eq. (\ref{eqa5}) can be reduced using the known relation
\begin{equation*}
    \arctanh(x) = \frac{1}{2} \ln \left(\frac{1+x}{1-x}\right),
\end{equation*}in which case, we obtain the following
\begin{equation}
    \label{eqa6}
    \rho_x(a; \m)=\rho_{x 0}
    \left[1-\frac{\beta}{2\m'-(1+b^2)}
    \left(\frac{1-y}{1+y}\right)
    \right]^{\frac{1}{1+\alpha}},
\end{equation}
where $\rho_{x0}$ is the present value of the
dark fluid, given by
\begin{equation}
    \label{rox0}\rho_{x0}^{1+\alpha}=\frac{B}{2\m'}(2\m'-(1+b^2)),
\end{equation}
and
\begin{equation}
    \label{eqa7}
    y\equiv \left(\frac{a_0}{a}\right)^ {3\beta(1+\alpha) }
    = \left(1+z\right)^ {3\beta(1+\alpha) }.
\end{equation}

\subsection{Dark matter density}

{Once the function $\rho_x(a, \m)$ has
been obtained, we will proceed to calculate the 
function $\rho_m(a, \m)$ from Eq. (\ref{eq:densmat}) 
which, by considering that $p_m=0$,
can be rewritten as
\begin{equation}
\label{eqrhm1}
    \frac{{\rm d}}{{\rm d}a}\left(\frac{a^3}{3} \rho_m\right)=b^2 a^2 \rho_x(a, \m), 
\end{equation}such that 
\begin{equation}
    \label{eqrhm}
    \rho_m(a, \m)=\rho_{m0}\left(\frac{a_0}{a}\right)^3 +\frac{3b^2}{a^3} \int_{a_0}^a \xi^2 \rho_x(\xi; \m) {\rm d}\xi.
\end{equation}
}
{Therefore, using Eq. (\ref{eqa4}) for
$\m=1$ we obtain
\begin{eqnarray}
    \label{eqrhom1}
    \rho_m(a, 1)&=&\left[
    \rho_{m0}-b^2\left(\frac{B}{1+b^2}\right)^{\frac{1}{1+\alpha}}\,_{2}F_1\left(c, \frac{c}{1+b^2}; 1+\frac{c}{1+b^2}; 1-\left(\frac{1+b^2}{B}\right)\rho_{x0}^{1+\alpha}\right)\right]\left(\frac{a_0}{a}\right)^3\\ \nonumber
    &&+ 
    b^2\left(\frac{B}{1+b^2}\right)^{\frac{1}{1+\alpha}}\,_{2}F_1\left(c, \frac{c}{1+b^2}; 1+\frac{c}{1+b^2}; 1-\left(\frac{1+b^2}{B}\right)\rho_{x}^{1+\alpha}(a, 1)\right),
\end{eqnarray}
where $c=-(1+\alpha)^{-1}$, and
$\,_2F_1(c_1, c_2; c_3; z)$ is the 
Gaussian hypergeometric function,
while for $0\leq \m <1$ we use Eq. (\ref{eqa6})
to obtain
\begin{eqnarray}
 \label{eqrhom2} 
 \rho_{m}(a, \m)&=&\left[\rho_{m0}-b^2 (1+\kappa)^{1+\alpha}\,F_1\left(-\frac{c}{\beta}; -c, c; 1-\frac{c}{\beta}; -1,-\frac{1-\kappa}{1+\kappa}\right) \right]\left(\frac{a_0}{a}\right)^3 \\ \nonumber
 &&+ b^2 (1+\kappa)^{1+\alpha} 
 \,F_1\left(-\frac{c}{\beta}; -c, c; 1-\frac{c}{\beta}; -\frac{1}{y},-\left(\frac{1-\kappa}{1+\kappa}\right)\frac{1}{y}\right),
\end{eqnarray}
where $F_1(c_1; c_2, c_3; c_4; z_1, z_2)$ is the two-parameter
Appell hypergeometric function, and
$$\kappa\equiv \frac{\beta}{2\m'-(1+b^2)}.$$

Note that the first Appell function in Eq. \eqref{eqrhom2} can be expressed in terms of the Gaussian hypergeometric function, as explained in the appendix \ref{app:A}. In Fig. \ref{fig:rhoxm}, we have depicted the $z$-profiles of $\rho_x$ and $\rho_m$, based on their expressions for $\m=1$ and $0\leq \m<1$. 
%
%

\begin{figure}[htbp!]
    \centering
    \includegraphics[width=0.8\textwidth]{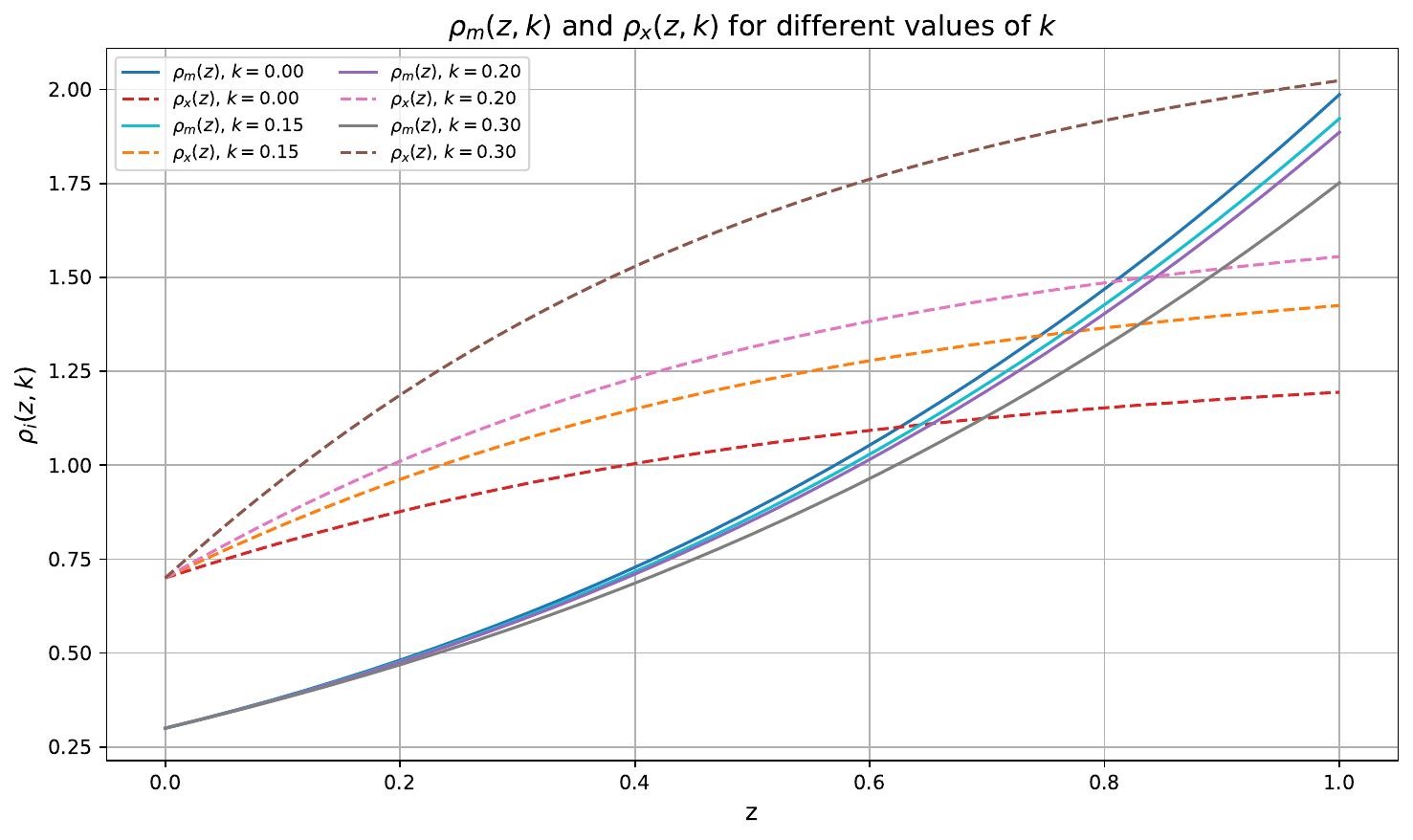}
    \caption{The $z$-profiles of the density functions $\rho_x$ and $\rho_m$, plotted for $\rho_{x0} = 0.7$, $\rho_{m0}=0.3$, $B=0.01$, $b=0.2$ and $\alpha=0.1$.}
    \label{fig:rhoxm}
\end{figure}
For completeness, the Hubble parameter $H(z)$ is constructed from the sum of the derived energy densities:
\begin{equation}
H(z) = \sqrt{\frac{1}{3}\bigl[\rho_{m}(z) + \rho_{x}(z)\bigr]},
\end{equation}
where $\rho_{x}$ and $\rho_{m}$ are the analytical solutions given in Eqs. (\ref{eqa6}) and (\ref{eqrhom2}) respectively, involving the Appell hypergeometric functions. According to Figure (\ref{fig:rhoxm}), the beginning of the dark energy dominance era, given by $\rho_{x}=\rho_{m}$, has a strong dependence on the parameter $\m$. In Fig. (\ref{fig:qterm}) we show the behavior of the interaction term using the same values as in Fig. (\ref{fig:rhoxm}) for the model parameters. The interaction term remains positive throughout cosmic evolution. It is worth noting that we present our results only for different values of the parameter $\m$, since the model is more sensitive to variations in this parameter than in the other, owing to its role in controlling deviations from the standard GCG model.

\begin{figure}[htbp!]
    \centering
    \includegraphics[width=0.8\textwidth]{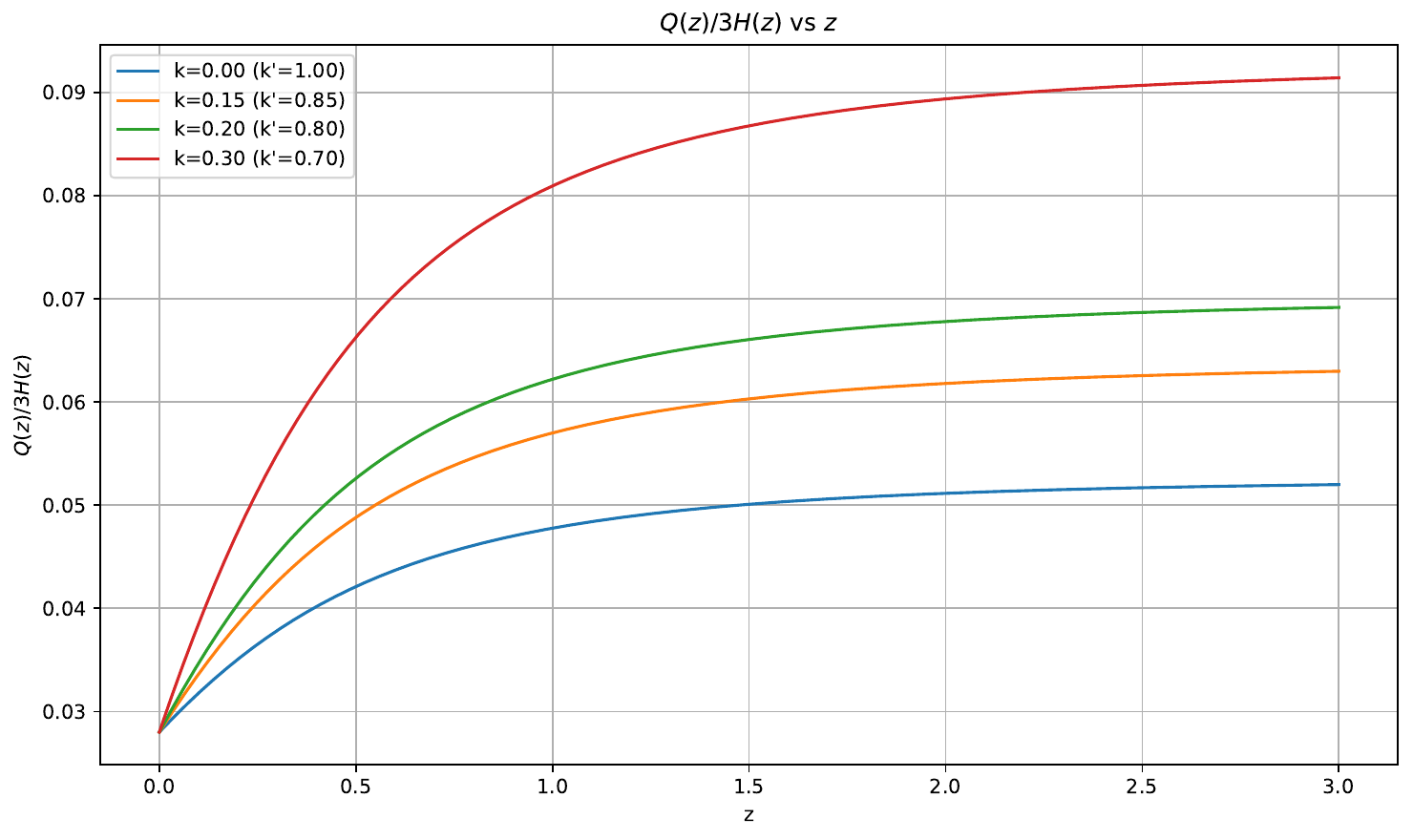}
    \caption{The $z$-profile for the $Q$-term, $Q = 3 H b^2 \rho_x$.}
    \label{fig:qterm}
\end{figure}

\begin{figure}[h!]
    \centering
    \includegraphics[width=0.8\linewidth]{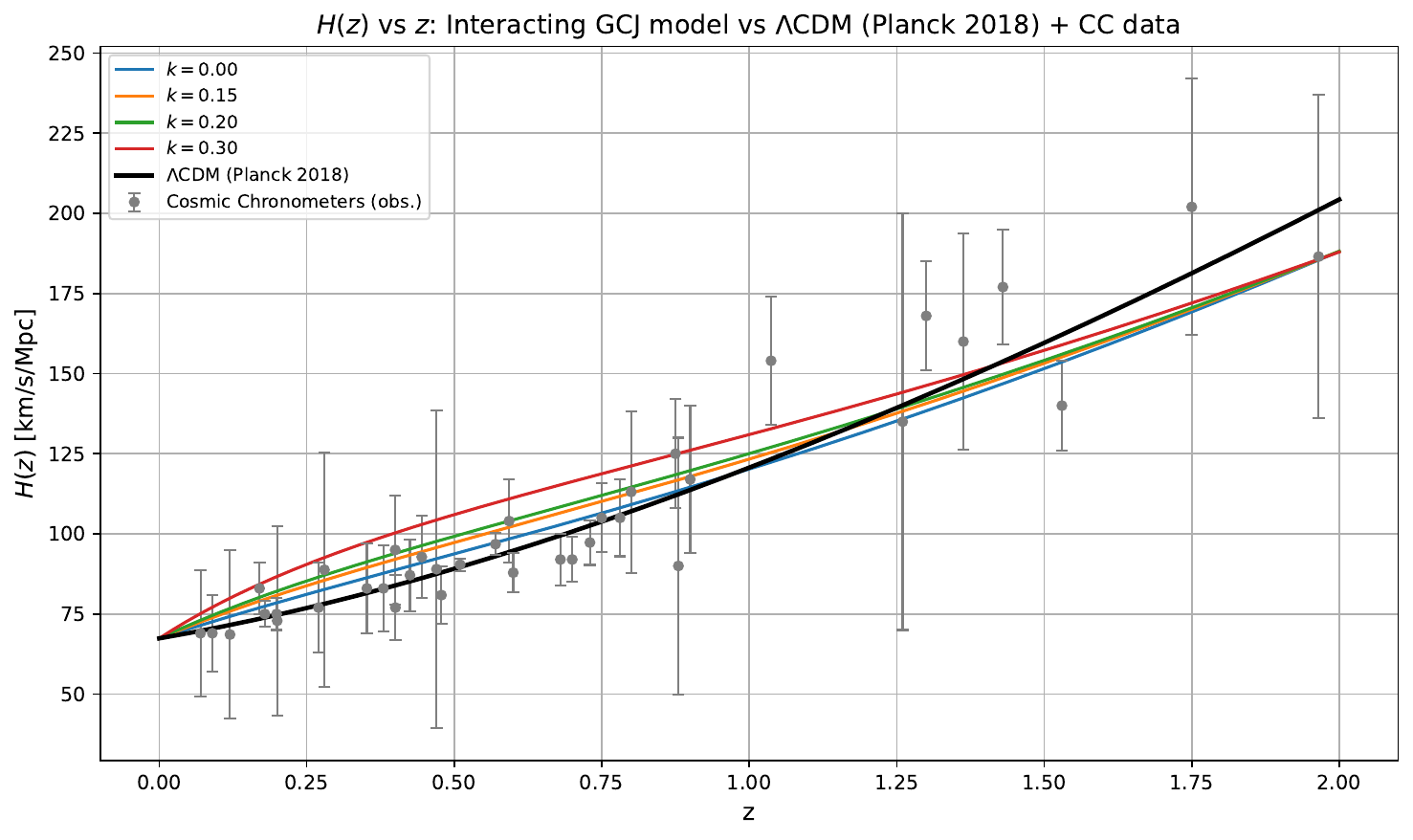}
    \caption{Hubble parameter as function of the redshift using the energy densities $\rho_{m}$ and $\rho_{x}$. We have considered the Planck 2018 results for the values of the cosmological parameters involved \cite{2020}.}
    \label{fig:HubbleP}
\end{figure}
Figure (\ref{fig:HubbleP}) presents the evolution of the Hubble parameter, $H(z)$, as a function of redshift for the interacting model compared to the standard $\Lambda$CDM model framework and the latest compilation of Hubble parameter measurements as a function of redshift obtained with cosmic chronometers; see Chapter 15 of reference \cite{di2023hubble}. The plot demonstrates that as the parameter $\m$ increases, the predicted value of $H(z)$ at low redshifts ($z \leq 1.0$) increases systematically over that of the $\Lambda$CDM model. This deviation could result in a higher value for the Hubble constant, $H_{0}$, in a full statistical analysis offering a potential late-time solution to the $H_{0}$ tension. Importantly, the figure shows that the predictions for the interacting model, particularly for $\m = 0.15, \m= 0.20$, and $\m = 0.30$, remain consistent with the cosmic chronometer data within their current observational uncertainties, validating the model's viability against direct measurements of the universe's expansion history. It is worth mentioning that the values used for the parameters $B, \alpha$ and $\m$ are also consistent with the bounds imposed by black hole physics \cite{Fathi_2024}.

\section{Thermodynamics behavior of the interacting scenario}\label{sec:thermo} 
In order to study the thermodynamic scenario of this cosmological model, we write equations (\ref{eq:densmat}) and (\ref{eq:densgcj}) in the following form
\begin{eqnarray}
   & \dot{\rho}_{m}+3H\rho_{m}(1+\omega_{\mathrm{eff},m})=0,\label{eq:ceffm}\\
   & \dot{\rho}_{x}+3H\rho_{x}(1+\omega_{\mathrm{eff},x})=0, \label{eq:ceffx}
\end{eqnarray}
where we have defined the effective parameter state for each component as follows
\begin{eqnarray}
    \omega_{\mathrm{eff},m}&:=& -\frac{Q}{3H\rho_{m}}, \label{eq:effm}\\
    \omega_{\mathrm{eff},x}&:=& \frac{p_{x}}{\rho_{x}}+\frac{Q}{3H\rho_{x}},\label{eq:effx}
\end{eqnarray}
where the matter component satisfies a cold dark matter equation of state, that is, it behaves as a pressureless fluid whose effective description exhibits a dynamical evolution \cite{M_B_Gavela_2009}. On the other hand, the pressure for dark energy is given as (\ref{eq:prro2}). Notice that we will obtain deviations from the standard case for cold dark matter due to the presence of the interaction term, $Q$. For $Q=0$, the parameter state of the dark matter sector is zero, and the barotropic case is recovered for dark energy $\omega_{x}=p_{x}/\rho_{x}$. In Fig. (\ref{fig:gamma_x}), we present the evolution of $\gamma_{x}\equiv 1+\omega_{x}$ for various choices of the model parameters.\\
\begin{figure}[h!]
    \centering
    \includegraphics[width=0.8\linewidth]{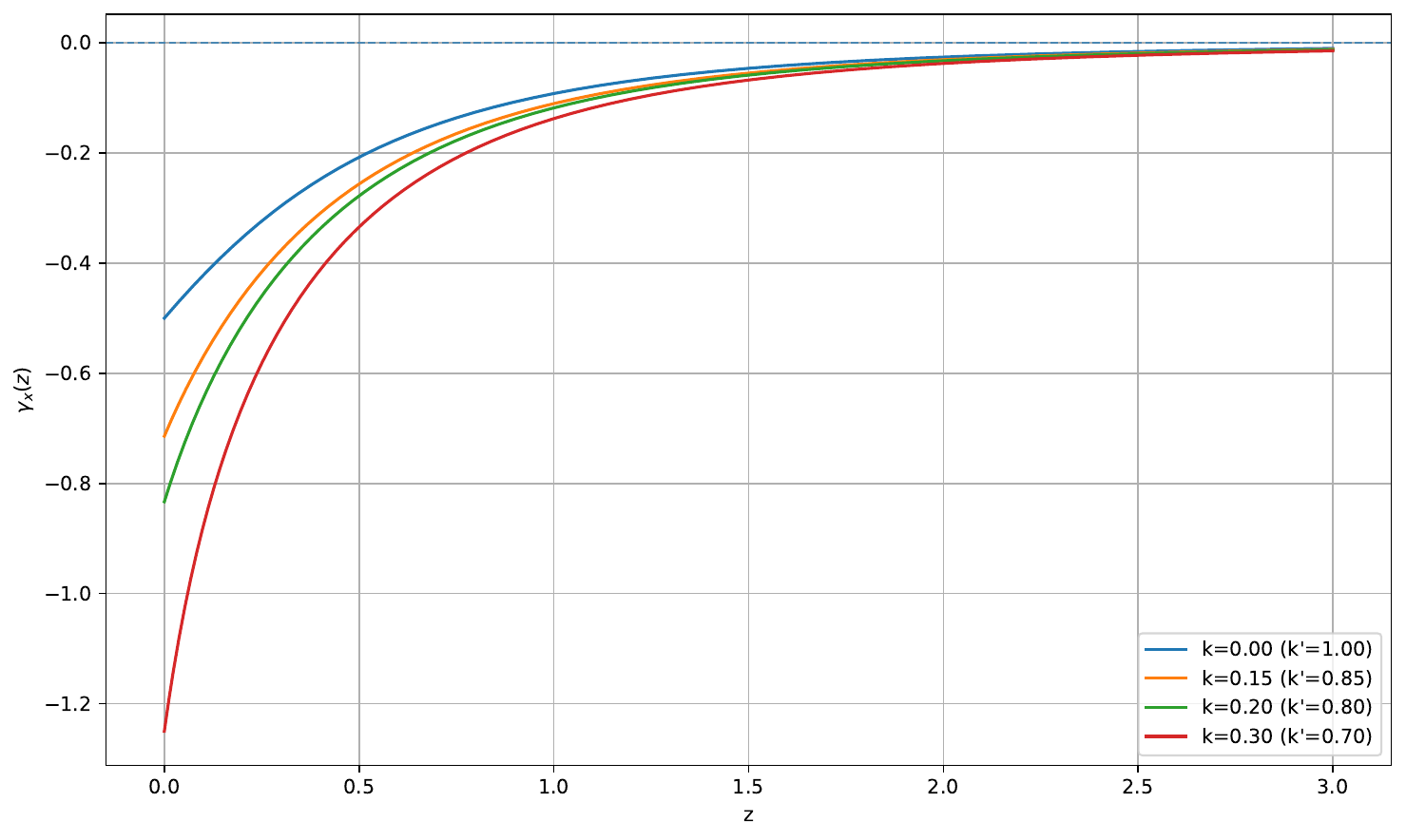}
    \caption{Behavior of $\gamma_{x}$ as function of the cosmological redshift.}
    \label{fig:gamma_x}
\end{figure}
The conservation equations given in the form (\ref{eq:ceffm}) and (\ref{eq:ceffx}) allow us to implement the effective temperature method developed in \cite{Cardenas_2019} for interacting fluids, which generalizes the single fluid description provided in Ref. \cite{maartens1996causalthermodynamicsrelativity}, yielding the following 
\begin{equation}
    T_{i}(z) = T_{i,0}\exp \left[\int^{z}_{0}\omega_{\mathrm{eff},i}(x)d\ln (1+x) \right], \label{eq:temps}
\end{equation}
for the temperature of both components of the dark sector, where $i=m,x$ and the change of variable from cosmic time to redshift, adopting the usual transformation $(1+z)=a(t)^{-1}$ was considered. Taking into account $Q = 3 H b^2 \rho_x$ and inserting equations (\ref{eq:effm}) and (\ref{eq:effx}), we solve the integral (\ref{eq:temps}) numerically for the temperatures of both sectors using the energy densities obtained before. From now on, we will consider the same values as those in the previous section for the parameters of the model. We show the behavior of the temperatures in the plot (\ref{fig:temps}) in terms of the cosmological redshift. The upper panel corresponds to the dark matter sector, and the lower panel corresponds to dark energy. Notice that both temperatures are positive, which is a signal of stability for thermodynamic systems and exhibit the typical behavior described in the interaction scheme. Additionally, because of the existence of energy interchange between the components, the temperature of the dark energy sector deviates from the value of zero, which is the value predicted by the standard cosmological scenario for dark energy. However, as stated by the third law of thermodynamics, this temperature value is not physically reachable \cite{callen2006thermodynamics}. In some sense, the temperature of the dark energy sector becomes relevant as dark energy becomes dominant in cosmic evolution. Additionally, for the dark matter sector, the behavior of the temperature differs from the constant value predicted by the standard scenario.\\
\begin{figure}[htbp!]
    \centering
    \includegraphics[width=0.8\textwidth]{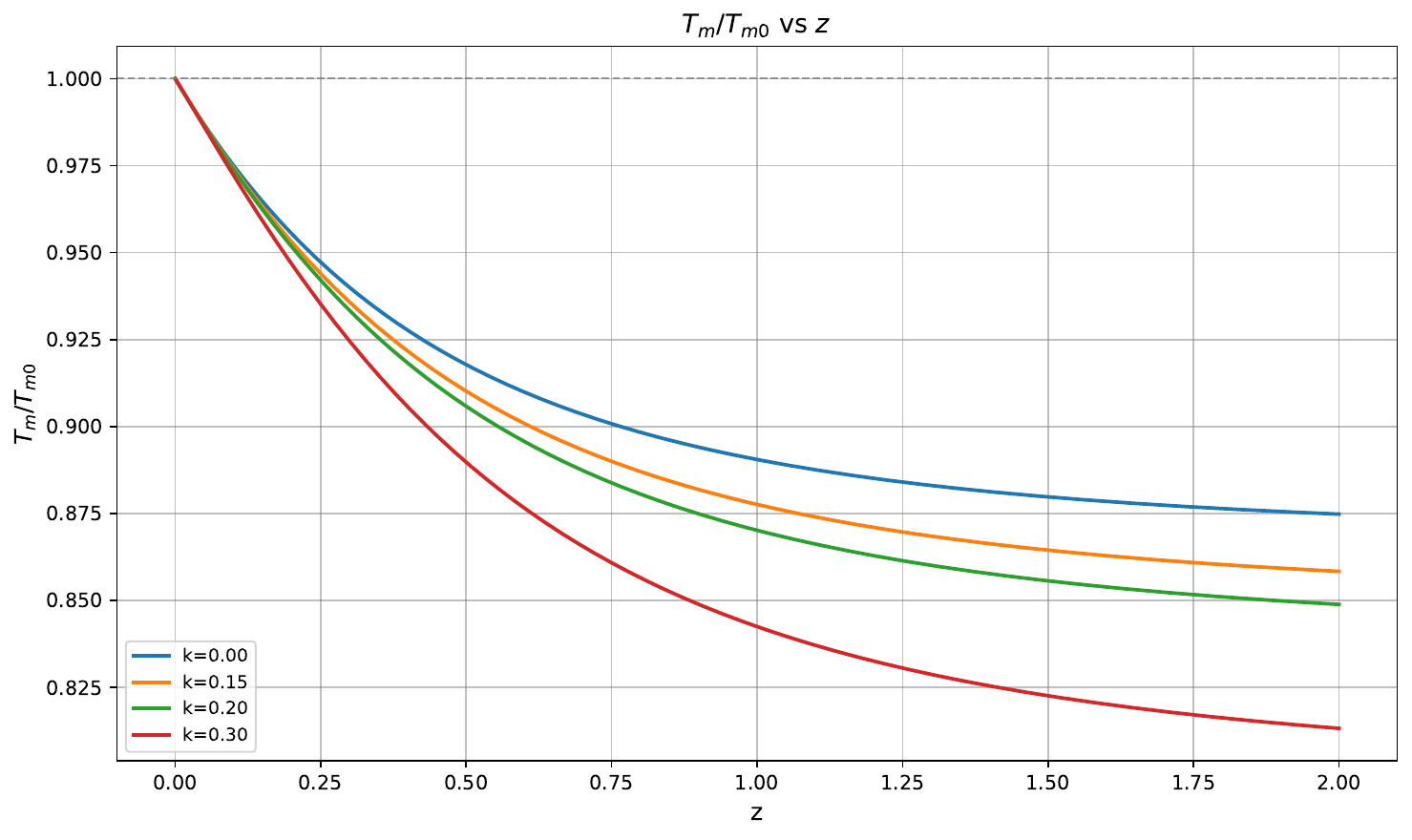}\\
   \includegraphics[width=0.8\textwidth]{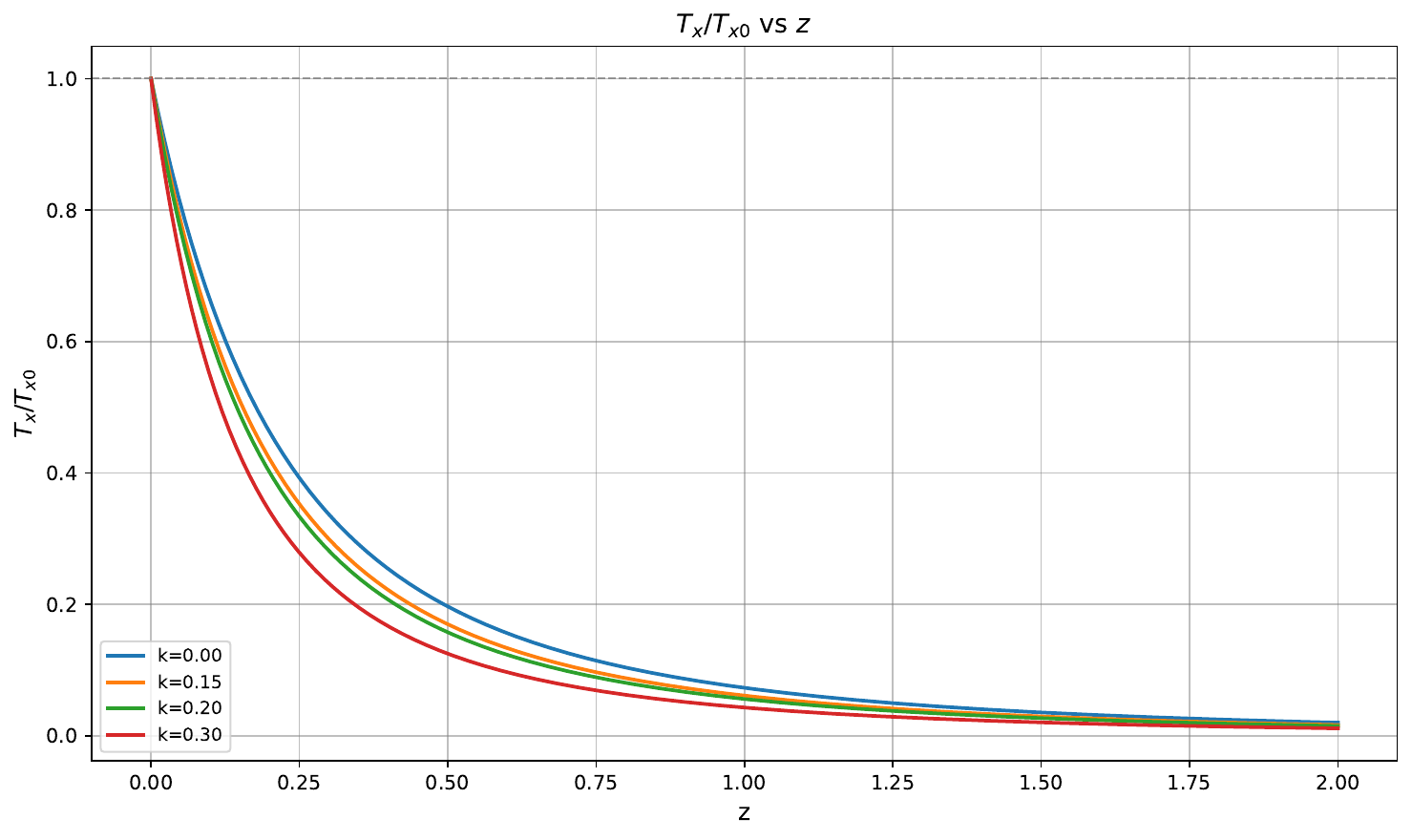}
    \caption{Temperatures of the dark sector as function of the cosmological redshift.}
    \label{fig:temps}
\end{figure} 

\begin{figure}[htbp!]
    \centering
    \includegraphics[width=0.8\textwidth]{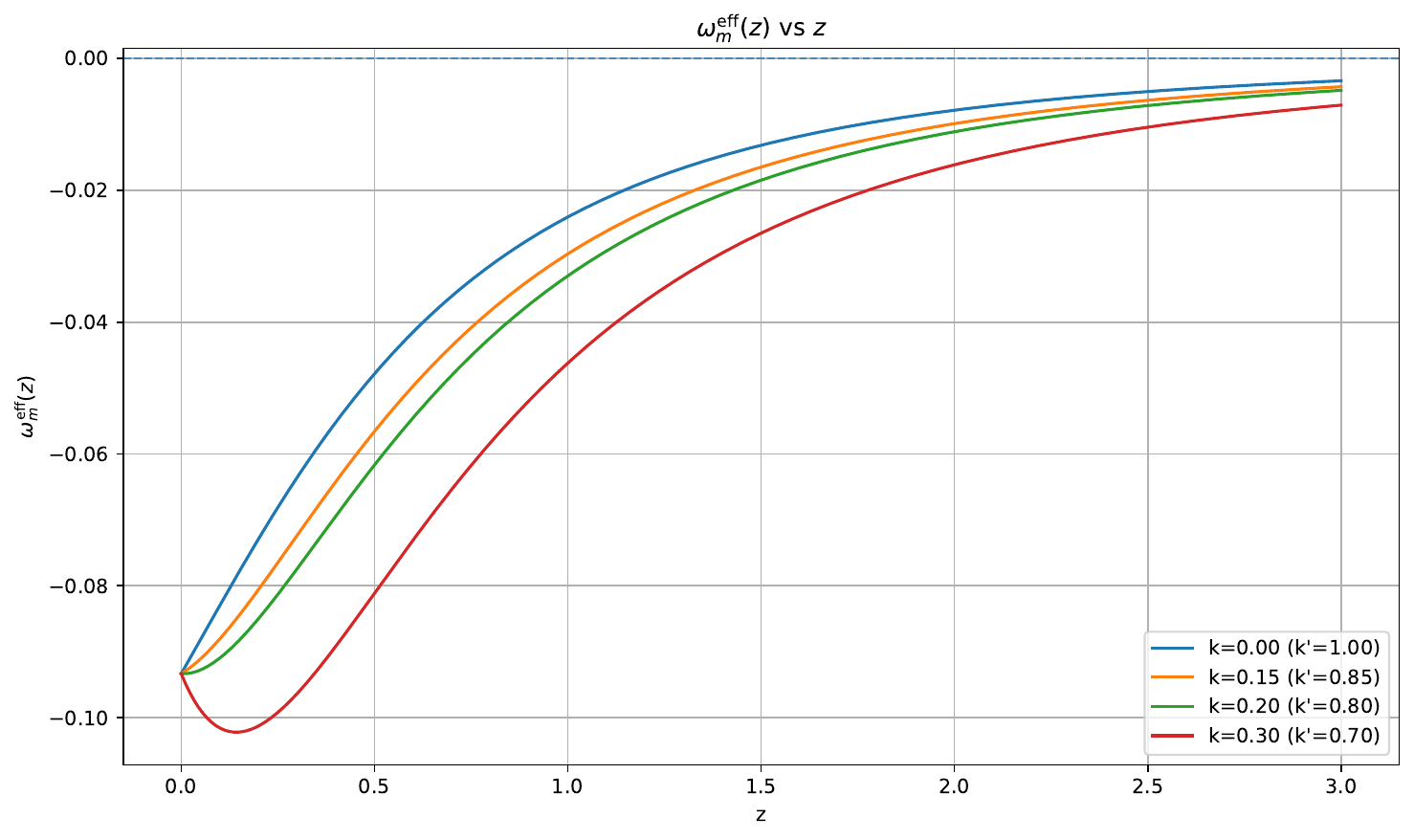}\\
   \includegraphics[width=0.8\textwidth]{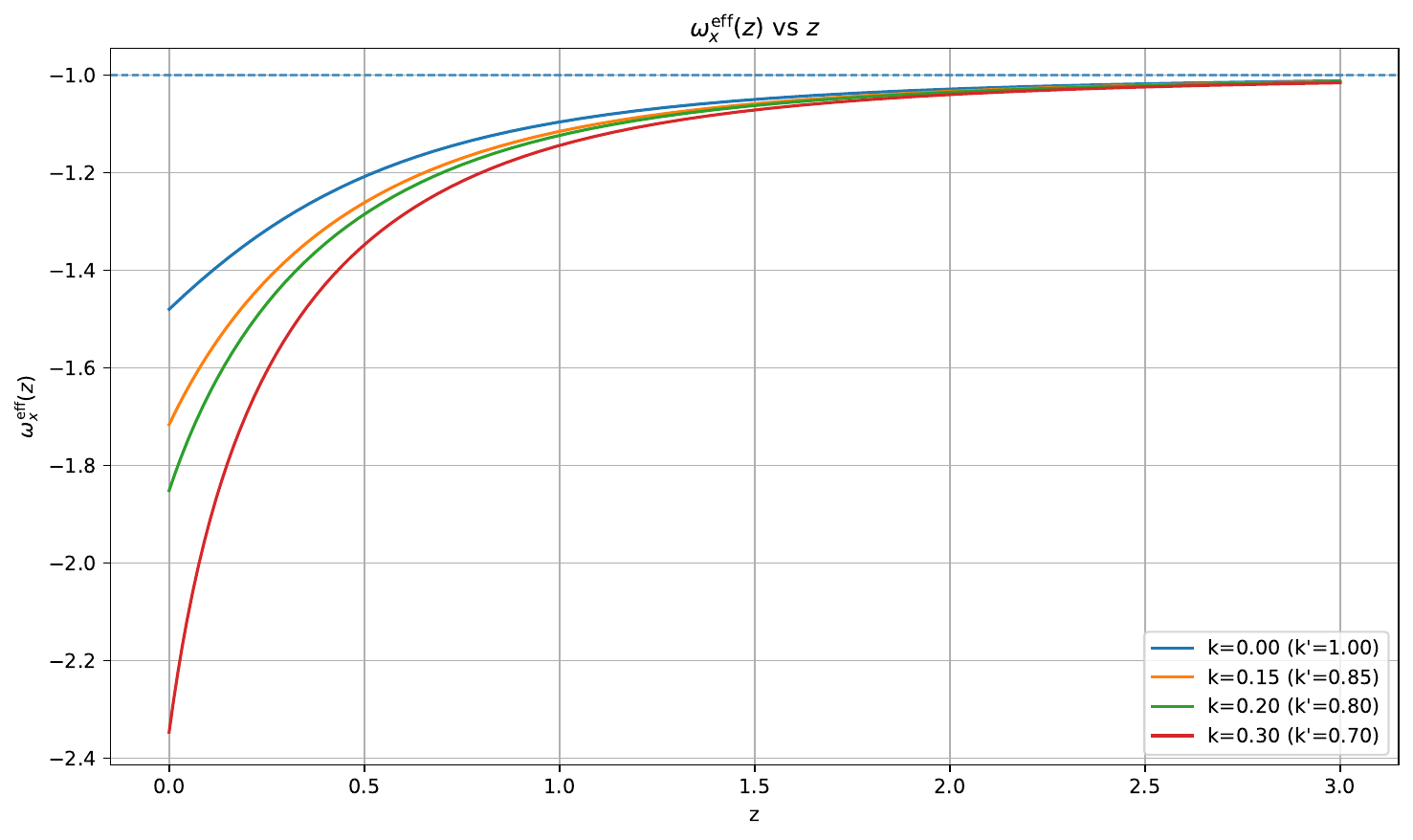}
    \caption{Effective parameter of state for the matter and the dark sector as functions of the cosmological redshift.}
    \label{fig:omegas}
\end{figure}
The behavior observed for the temperatures of the constituents of the dark sector corresponds to the behavior of their effective parameter states. If we compare the plots given in figures (\ref{fig:temps}) and (\ref{fig:omegas}), we observe that for $z\geq 2$, the temperature of dark matter exhibits a slowly increasing behavior. In this zone, its parameter state is close to zero value. For $0 \leq z \leq 2$, the temperature tends to reach its maximum value, while the effective parameter state reaches its minimum value. Something similar occurs for dark energy, whereas the effective parameter state is close to $-1$, which resembles the behavior of a cosmological constant; its temperature is very close to zero. Once the effective parameter state departs from the value $-1$, the temperature increases, reaching its maximum value at the present time $(z=0)$. From the plots presented in Fig. (\ref{fig:omegas}) for the dark matter sector, we notice slight deviations from zero in the effective parameter state, indicating that, at the effective level, the dark matter sector does not conform strictly to the standard {\it cold} dark matter description. These results are in agreement with other works, where a non-zero EoS for dark matter ($\omega_{\mathrm{m}} > 0$ or $\omega_{\mathrm{m}} < 0$) is validated by using different datasets \cite{dm1, dm2}. This kind of scenario generalizes the description of dark matter, and the consideration of $\omega_{m}\neq 0$ has been studied extensively; see, for instance, the references given in \cite{Hu_1998, Kopp_2018, Kopp_2016}, see also \cite{Ju_rez_Jim_nez_2025} where something similar occurs in the context of unstable dark matter. 

It is important to mention that for the dark energy case, we observe a transition from a cosmological constant-like scenario to a phantom regime. In all cases, the phantom crossing $\omega < -1$ takes place in the past $(z>0)$. This is an interesting scenario, as a possible explanation for a better interpretation of the latest results released by the DESI collaboration requires early phantom dark energy; see, for example, \cite{Cort_s_2024}. On the other hand, we observe that, at an effective level, the phantom regime emerges without the need to impose a negative energy density at some stage of cosmic evolution, as required in other interacting scenarios \cite{Quartin_2008}. Notice that this phantom scenario also maintains the positivity of its temperature, as required by standard thermodynamics, to have well-defined physical quantities. Although a negative temperature may be allowed, it could lead to a pathological definition of entropy, see Ref. \cite{Cruz_2023}, where these aspects of the phantom regime were discussed.

Now, we consider the second law of thermodynamics,
\begin{equation}
    TdS = dU+pdV, \label{eq:second}
\end{equation}
where $V$ is the Hubble volume defined as $V=V_{0}(a/a_{0})^{3} = V_{0}(1+z)^{-3}$, and $U$ is the internal energy $U=\rho V$. If we consider the continuity equations in the form given by (\ref{eq:densmat}) and (\ref{eq:densgcj}), we can write the following expression in terms of the redshift for the second law (\ref{eq:second})
\begin{equation}
    -\frac{T_{m}(z)}{V(z)}\frac{dS_{m}(z)}{dz} = \frac{Q}{H_{0}E(z)(1+z)} = -\frac{T_{x}(z)}{V(z)}\frac{dS_{x}(z)}{dz},
\end{equation}
where $E(z)$ is the normalized Hubble parameter, $E(z) = H(z)/H_{0}$, with $H_{0}$ being the value of the Hubble parameter at the present time, $H_{0} = H(z=0)$. Notice that the entropy production per component is not constant, the adiabatic case ($S=\mbox{constant}$) is recovered for $Q=0$. From our latter result, we can write
\begin{equation}
    \frac{d}{dz}\left(S_{x}(z)+S_{m}(z) \right)= \frac{Q}{H_{0}E(z)(1+z)}\left(\frac{1}{T_{x}(z)} - \frac{1}{T_{m}(z)}\right) = \frac{Q}{H_{0}E(z)(1+z)}\frac{1}{T_{x}(z)}\left(1-\frac{T_{x}(z)}{T_{m}(z)}\right), \label{eq:second2}
\end{equation}
Then, we observe that the fulfillment of the second law of thermodynamics, $dS_{T} \geq 0$, depends on the sign of the interaction term and the behavior of the temperatures of the dark sector, which were obtained numerically. Due to the change of variable from cosmic time to cosmological redshift, the second law: $\dot{S}_{T} \geq 0$ turns into $S'_{T} \leq 0$, where the prime denotes derivative w.r.t. the cosmological redshift. In order to fulfill the second law of thermodynamics, for $Q>0$, we must have $T_{x} > T_{m}$, as also obtained in \cite{Cardenas_2019}. Therefore, at present time the value of the derivative (\ref{eq:second2}) is negative for $T_{m,0} < T_{x,0}$\footnote{This contrasts with the $\Lambda$CDM model, where the matter sector has a constant temperature and the dark energy sector is characterized by null temperature, which constitutes a violation of the third law of thermodynamics.}. Taking into account this latter condition together with the analytical solutions for the densities and the numerical solutions obtained previously for the temperatures in the Eq. (\ref{eq:second2}), we obtain the production of the total entropy setting $V_{0}=1$. As can be seen in Fig. (\ref{fig:entropy}), the cosmic expansion evolves from a stage near to the equilibrium condition ($S'_{T}\simeq 0$) in the past and tends to a non-adiabatic expansion ($S'_{T}\leq 0$) at late times. As stated in \cite{Maartens_1997}, a viable thermodynamic description of the cosmic expansion requires the inclusion of deviations from the equilibrium condition within the cosmic fluid, see also \cite{ Aguilar-Pérez_2022}.

As discussed in \cite{GARCIABELLIDO2021100892}, the existence of entropy gradients ({\it entropic forces}) at some stage of cosmic evolution could be responsible for the observed acceleration at late times. Notice that in the interaction scenario such {\it forces} are mediated by the $Q$-term only.
\begin{figure}[htbp!]
    \centering
    \includegraphics[width=0.75\textwidth]{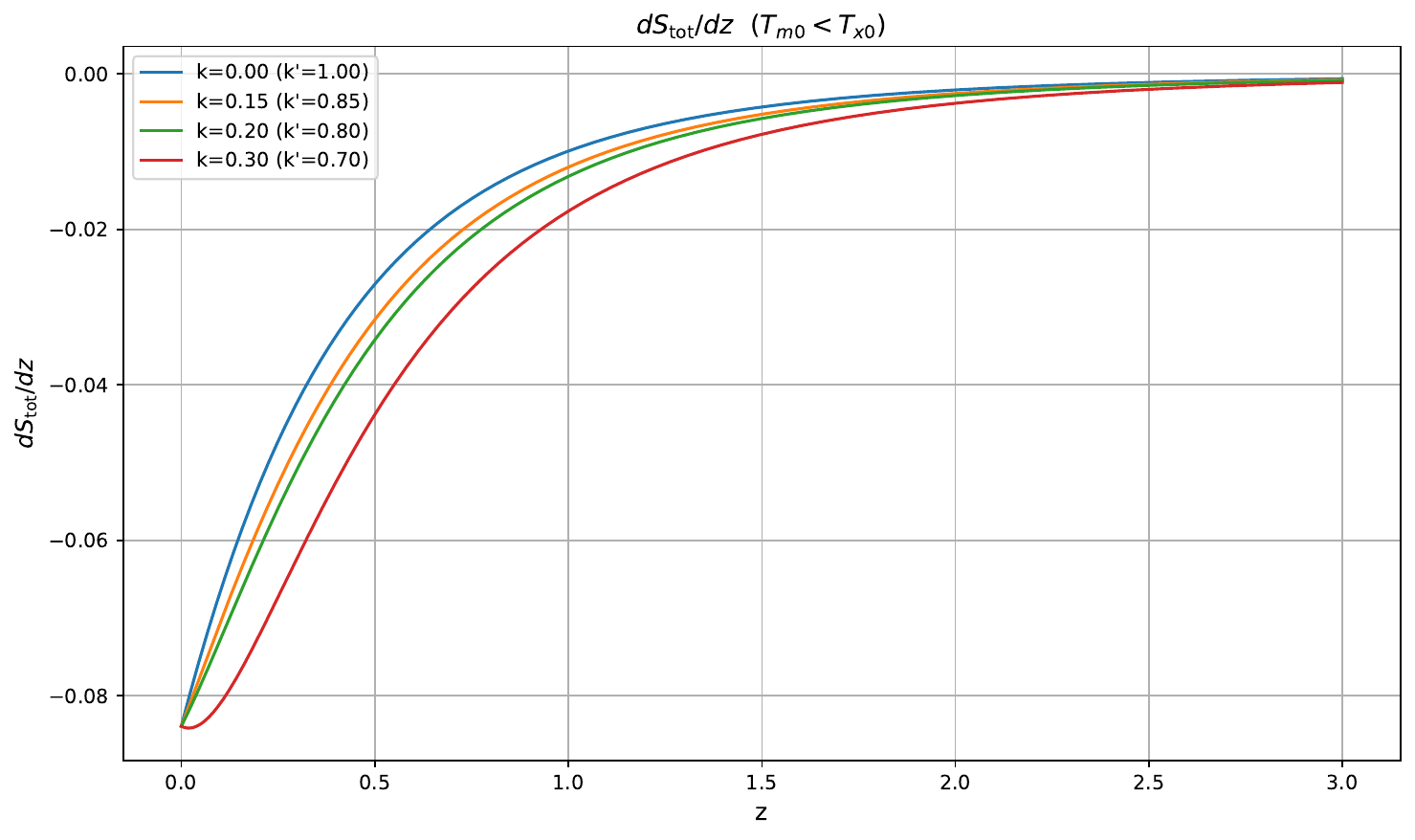} 
    \caption{Total entropy production in terms of the cosmological redshift.}
    \label{fig:entropy}
\end{figure}

Using the standard definitions of classical thermodynamics, we can write \cite{callen2006thermodynamics}
\begin{equation}
    C_{V} = \frac{\partial U}{\partial T}, \qquad C_{p} = \frac{\partial h}{\partial T},   
\end{equation}
and both expressions represent the specific heat at constant volume, $V$, and at constant pressure, $p$, respectively; $U$ and $h$ are the internal energy of the system and its enthalpy, which are defined as follows $U=\rho V$ and $h=(\rho+p)V$. $T$ is the temperature of the fluid. Given that all quantities depend on the cosmological redshift, we can compute the specific heats as $C_{V}=(dU/dz)(dT/dz)^{-1}$ and $C_{p}=(dh/dz)(dT/dz)^{-1}$. Using again the analytical and numerical solutions as before, we show the behavior of the specific heats for the dark sector. Since $C_{V}$ and $C_{p}$ depend on temperature, we consider the condition that guarantees the second law of thermodynamics: $T_{m,0} < T_{x,0}$. 

In Figure (\ref{fig:cpcvmatter}) we depict the behavior of the specific heats for the matter sector, the upper panel represents $C_{p}$, while the lower panel corresponds to $C_{V}$. Note that $C_{V} < 0$ and satisfies the limit $C_{V}(z\rightarrow 0) \simeq 0$ whereas $C_{p}$ is positive and reaches its maximum value around $z\simeq 0$, the difference between them obeys $C_{p}-C_{V} > 0$, which is a signal of thermodynamic stability. 
\begin{figure}[htbp!]
    \centering
    \includegraphics[width=0.8\textwidth]{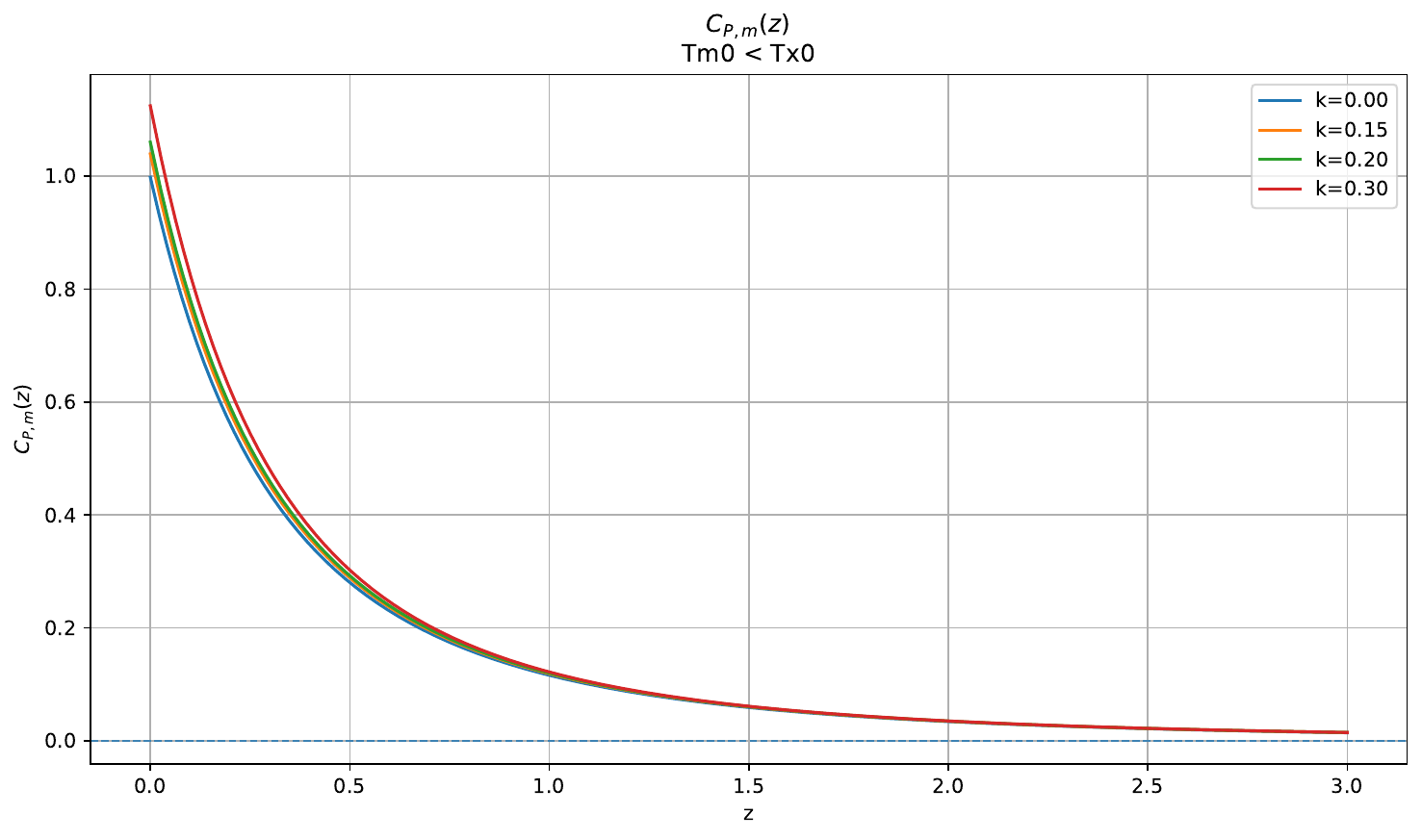}
    \includegraphics[width=0.8\textwidth]{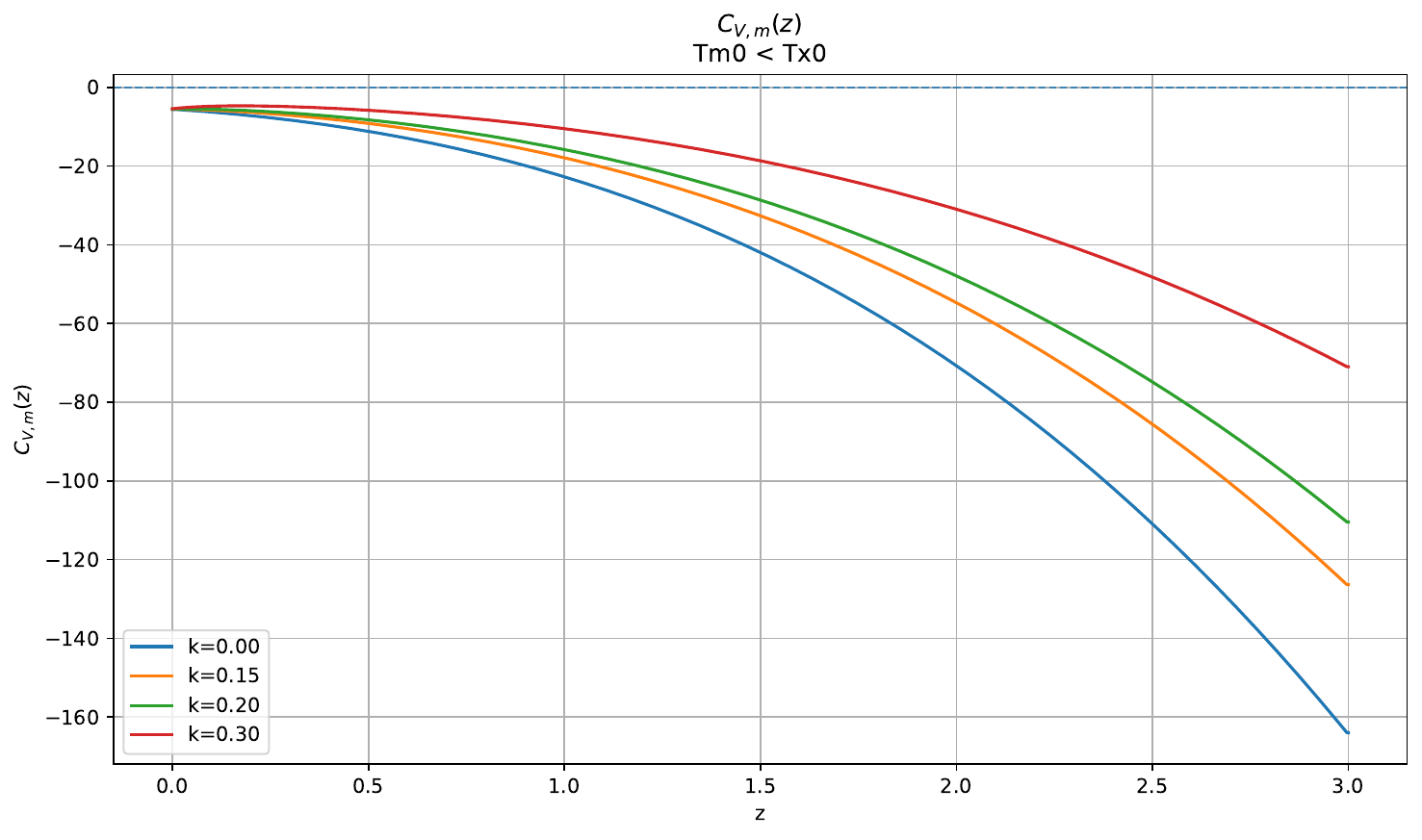}
    \caption{Specific heats, $C_{p}$ and $C_{V}$ for matter sector as function of $z$.}
    \label{fig:cpcvmatter}
\end{figure}
For the dark energy sector, the specific heats are presented in Figure~(\ref{fig:cpcvdark}). In the case of $\m = 0$, the heat capacity at constant pressure is positive ($C_p > 0$) and the heat capacity at constant volume is negative ($C_V < 0$) throughout the cosmic evolution, except at the present epoch ($z=0$) where $C_p = 0$. For non-zero values of $\m$, we observe that $C_p$ undergoes a sign change at low redshifts. Consequently, at late times ($z \simeq 0$), the condition $C_p - C_V > 0$ holds for $\m = 0$, whereas for $\m \neq 0$, this difference becomes negative.

These results reveal a significant contrast between the two cosmic components. The matter sector, characterized by $C_p > 0$ and $C_V < 0$, remains thermodynamically stable. However, the dark energy sector exhibits a phase transition in the late-time, as indicated by the sign change in $C_p$ \cite{Lepe_2016, PhysRevD.92.123511}. From a thermodynamic perspective, this instability corresponds to the model's transition from a quintessence-like to a phantom-like scenario, which is consistent with the previously observed behavior of the effective EoS for dark energy.

For the specific case of $\m = 0$ at $z=0$, the specific heats of the dark energy sector ($C_p = 0$ and $C_V < 0$) align with the predictions of the standard $\Lambda$CDM model. As noted by Luongo \& Quevedo, a null $C_p$ can characterize the transition of the universe from a decelerated to an accelerated phase of expansion~\cite{luongo_cosmographic_2013}. Additionally, for $\m \neq 0$ the specific heats for dark energy have the same sign at $z=0$, which is consistent with an accelerated universe, as established in Ref. \cite{Saha_2025}

Finally, we would like to comment that in our description the conditions obtained: $C_p - C_V < 0$ or $C_p - C_V > 0$ indicate a significant deviation from the behavior of an ideal gas, for which this difference is always positive and equal to the ideal gas constant, usually denoted by $R$. This result suggests that the cosmological fluids are governed by long-range gravitational interactions, requiring a thermodynamic framework that extends beyond standard statistical mechanics.

\begin{figure}[htbp!]
    \centering
    \includegraphics[width=0.8\textwidth]{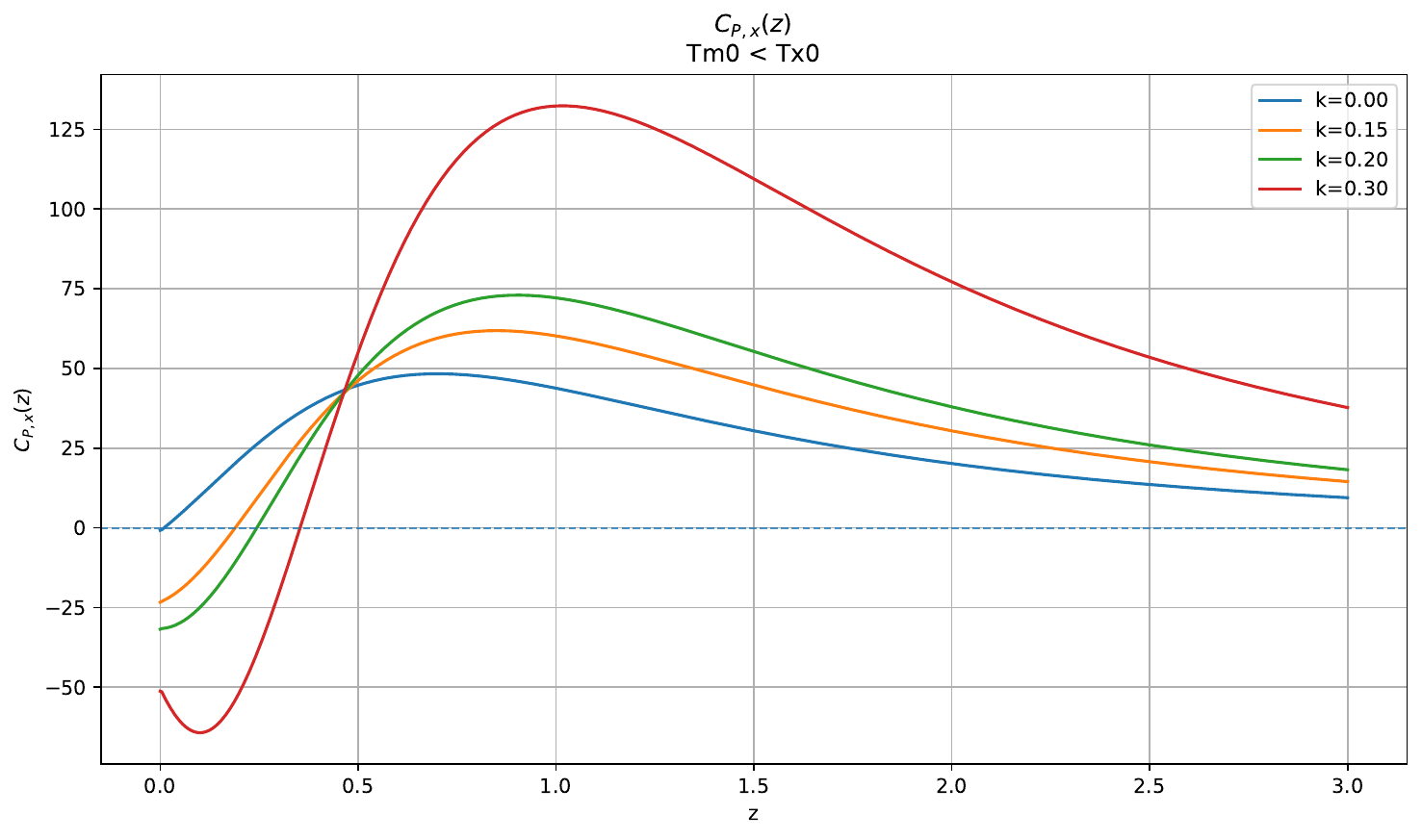}
    \includegraphics[width=0.8\textwidth]{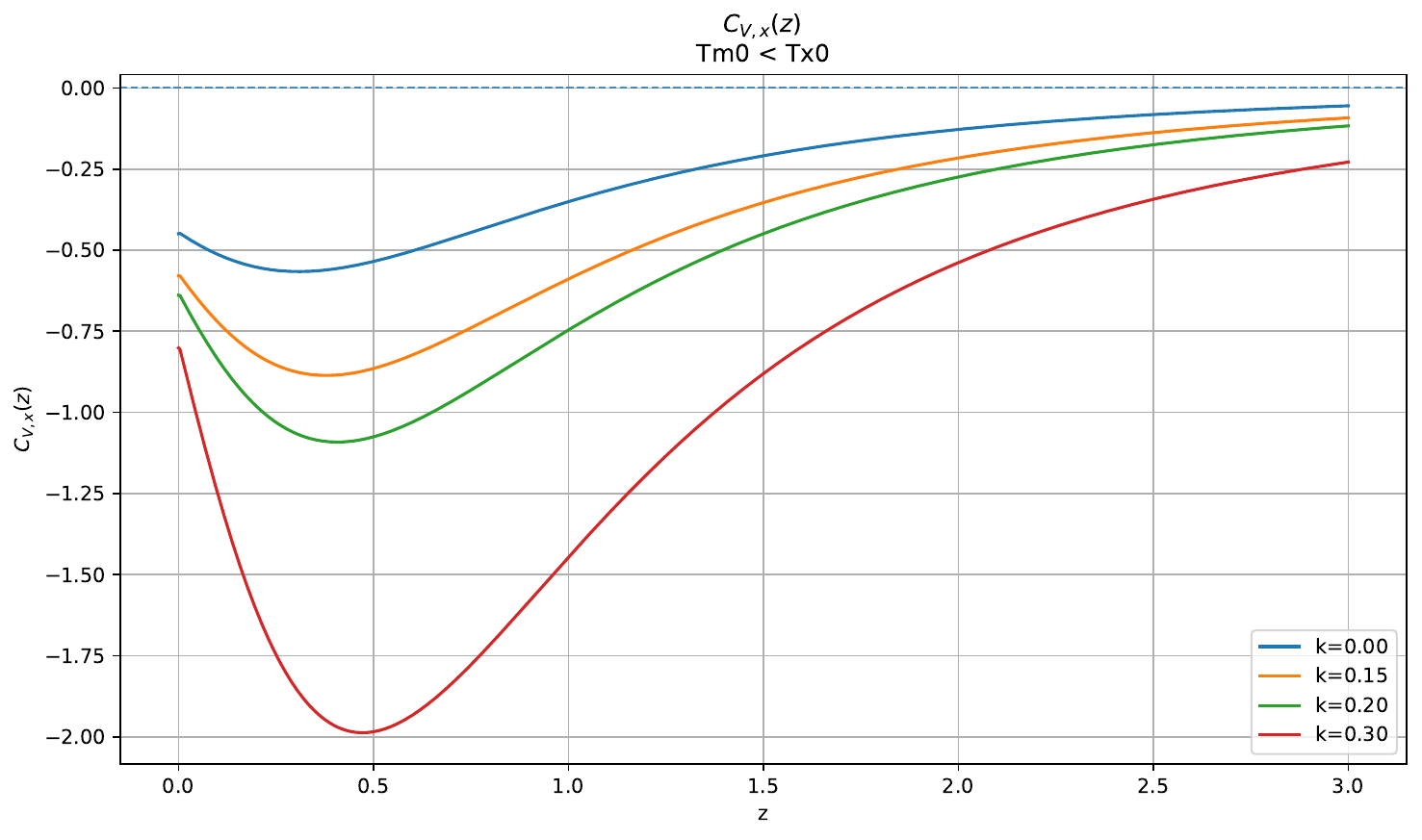}
    \caption{Specific heats, $C_{p}$ and $C_{V}$ for the dark energy sector as function of $z$.}
    \label{fig:cpcvdark}
\end{figure}

\section{Summary and conclusions}\label{sec:conclusions}

In this work we introduced and analyzed a cosmological scenario in which the dark sector features an explicit interaction between pressureless dark matter and a GCJG, the latter acting as a unified dark energy component. The interaction was modeled through the linear coupling $Q = 3 H b^{2} \rho_{x}$, allowing energy transfer from dark energy to dark matter at the background level. Within this framework we derived \emph{exact} analytical solutions for the energy densities of both dark components. Although their explicit form involves Appell hypergeometric functions, these solutions provide a mathematically controlled basis for exploring the cosmological implications of GCJG scenarios with interactions. The adoption of a positive coupling parameter, $b^{2}>0$, was selected based on both thermodynamic and observational considerations. This choice ensures a unidirectional energy flow favored by the second law of thermodynamics and recent statistical evidence. Moreover, requiring this parameter to remain positive prevents early-time, non-adiabatic instabilities and alleviates the cosmological coincidence problem that arises for negative coupling constants.

A key component of the analysis involved evaluating the thermodynamic feasibility of the model throughout the parameter space spanned by $B$, $\m$, and $\alpha$. Both dark matter and dark energy were shown to maintain positive temperatures throughout cosmic evolution, ensuring that the interacting system does not develop thermodynamic pathologies. We further confirmed that the entropy production complies with the second law of thermodynamics, as ensured by the condition $T_{m} < T_{x}$, which remains valid across our model for the parameters examined in the numerical solutions.

A particularly interesting feature arises from the dynamical behavior of the EoS of GCJG. We found that the effective dark energy EoS naturally crosses into the phantom regime in the past, a behavior that may help
to interpret recent DESI indications of a mild phantom trend at intermediate redshifts, without requiring negative energy densities or violating thermodynamical principles. At late times, the effective state parameter of dark matter exhibits a small deviation from the cold matter value $\omega_{\mathrm
m}=0$, suggesting that the interaction induces a mild departure from perfect pressureless behavior. An open question is whether adopting alternative coupling prescriptions would modify this phenomenology. Previous studies (e.g.~\cite{dm2}) have shown that non-cold dark matter cannot generally solve the $H_{0}$ tension within non-interacting dark energy models, due to degeneracies between $\omega_{\mathrm m}$ and $H_{0}$. However, this conclusion
might change in the presence of dark sector interactions, and our model provides a useful setting in which to explore this possibility.

The thermodynamic analysis of the specific heats shows that while the dark matter sector remains thermally stable, the GCJG component undergoes a late-time phase transition that aligns with its transition toward a phantom-like regime. Such features highlight the rich phenomenology that emerges when nonlinear equations of state and dark-sector interactions coexist.

This work suggests several promising directions for further investigation. First, the analytical expressions derived here allow for a direct construction of the Hubble parameter, enabling a full observational analysis using CMB, BAO, SNIa, cosmic chronometers, and DESI-like data. This will allow for a precise
determination of the viable parameter space and a comparison with competing cosmological scenarios. Second, a natural extension is to study linear cosmological perturbations in the interacting GCJG model, including the growth rate of matter fluctuations and the imprint on CMB temperature and polarization spectra. This is essential for assessing whether nonlinear dark energy behavior
combined with interactions can address current cosmological tensions. Third, model selection techniques (e.g., AIC, BIC, Bayesian evidence) will be necessary to evaluate whether the increased complexity of the GCJG interaction is statistically justified. Finally, given the connection between the GCJG
EoS and Jacobi-type potentials, it would be interesting to explore potential links with scalar field realizations or modified gravity embeddings, which may yield additional theoretical insights.

In summary, the interacting GCJG model constitutes a thermodynamically viable and phenomenologically rich alternative to standard dark-sector descriptions. Its nonlinear EoS, combined with a controlled energy exchange, generates behaviors (such as phantom crossing, late-time phase transitions, and effective deviations from cold dark matter) that could have observable signatures. Future observational and theoretical work will be crucial in assessing the full potential of this framework to describe the dynamics of the dark
sector.

\begin{acknowledgments}
G.A.P. was supported by SECIHTI through the {\it Estancias Posdoctorales por México 2023(1)} program. G.A.P. and M.C. work was partially supported by S.N.I.I. (SECIHTI-M\'exico). J.R.V. is partially supported by the Centro de F\'isica Teórica de Valparaíso (CeFiTeV).
\end{acknowledgments}

\appendix

\section{Reduction formulae for the Appell hypergeometric functions}\label{app:A}

The two Appell functions in Eq. \eqref{eqrhom2} can be reduced as \cite{Erdelyi1953HTF,Slater1966,Luke1969,Exton1976,schneider_multiple_2013}
\begin{eqnarray}
F_1\left(-\frac{c}{\beta}; -c, c; 1-\frac{c}{\beta}; -1,-\frac{1-\kappa}{1+\kappa}\right) &=& (1+\kappa)^c\,F_1\left(1;-c,c;1-\frac{c}{\beta};\frac{1}{2},\frac{1-\kappa}{2}\right)\nonumber\\
&=& \left(\frac{1-\kappa}{1+\kappa}\right)^c\,_2F_1\left(1,1-\frac{c}{\beta};1-\frac{c}{\beta};\frac{2\kappa}{1+\kappa}\right).
    \label{eq:F1a_red}
\end{eqnarray}
In the same way, the second Appell function can be recast as
\begin{eqnarray}
&& F_1\left(-\frac{c}{\beta}; -c, c; 1-\frac{c}{\beta}; -\frac{1}{y},-\left(\frac{1-\kappa}{1+\kappa}\right)\frac{1}{y}\right) \nonumber\\
&&= \left(1+\frac{1}{y}\right)^{c/\beta}\left(1-\frac{1}{1+y}\right)^{c/\beta-1}\left(1-\frac{\tilde{\kappa}}{1+y}\right)^{-c}\, _2F_1\left(-\frac{c}{\beta},-c;1-\frac{c}{\beta};\frac{(1-\tilde{\kappa})(1+y)}{y(1+y-\tilde{\kappa})}\right),
\label{eq:F1b_red}
\end{eqnarray}
in which $\tilde{\kappa} = 2\kappa/(1+\kappa)$. 

\bibliographystyle{ieeetr}
\bibliography{biblio_v1}

\end{document}